\documentclass[%
 reprint,
superscriptaddress,
 amsmath,amssymb,
 aps,
prx,
floatfix,
]{revtex4-2}

\usepackage{graphicx}
\usepackage{dcolumn}
\usepackage{bm}

\usepackage{verbatim}
\usepackage{textgreek}

\usepackage[caption=false]{subfig}
\usepackage{braket}

\usepackage{xcolor}
\usepackage{lipsum}
\usepackage{hyperref}

\begin{document}

\preprint{APS/123-QED}

\title{Noise-induced quantum-circuit refrigeration}

\author{Heidi Kivijärvi}
\email{Corresponding author email: heidi.kivijarvi@aalto.fi}
\affiliation{QCD Labs, QTF Centre of Excellence, Department of Applied Physics, Aalto University, FI-00076 Aalto, Finland}
\author{Arto Viitanen}%
\affiliation{QCD Labs, QTF Centre of Excellence, Department of Applied Physics, Aalto University, FI-00076 Aalto, Finland}
\author{Timm Mörstedt}%
\affiliation{QCD Labs, QTF Centre of Excellence, Department of Applied Physics, Aalto University, FI-00076 Aalto, Finland}
\author{Mikko Möttönen}%
\affiliation{QCD Labs, QTF Centre of Excellence, Department of Applied Physics, Aalto University, FI-00076 Aalto, Finland}
\affiliation{VTT Technical Research Centre of Finland Ltd, QTF Centre of Excellence, P.O. Box 1000, FI-02044 VTT, Finland}

\date{December 5, 2024}

\begin{abstract}
We use a transmon qubit and its dispersively coupled readout resonator to measure the Fock state populations of another microwave resonator, to which we have attached a quantum-circuit refrigerator (QCR). First, we apply noise generated at room temperature to the resonator and show that such noise drive leads to a thermal distribution of the resonator Fock states. Subsequently, we detune the noise frequency band far away from the resonance condition and vary the power of the noise applied on the QCR. We observe that such artificial thermal noise may lead to major damping of a coherent state of the resonator. Importantly, we also demonstrate that the effective temperature of a thermal resonator state can be reduced from roughly 300~mK to 130~mK by the introduction of the artificial thermal noise. These observations pave the way for a purely thermally powered quantum-circuit refrigerator which may unlock the use of waste heat in resetting superconducting qubits in a quantum processor and in building autonomous quantum heat engines.
\end{abstract}

\maketitle

\section{\label{sec:intro}Introduction}

Superconducting quantum circuits~\cite{Blais2021} have demonstrated favorable properties in pursuit of practical-use quantum-information systems~\cite{Krantz2019, Huang2020, Kjaergaard2020, Barends2014, Gunyho2024, Sunada2022, Geerlings2013, Magnard2018, Arute2019, Kim2023}. Thus far, most initialization, quantum-gate, and readout protocols of superconducting qubits rely on the common practice of transmitting power from a room-temperature source through a set of attenuators to the millikelvin-temperature quantum circuit inside a dilution refrigerator~\cite{Ezratty2023, Krinner2019}. Consequently, the on-going major scale-up of these systems calls for novel solutions that ensure reasonable millikelvin power consumption without compromising the system performance~\cite{Almudever2017}. Natural candidates, yet lacking wide interest, are robust on-chip nanoelectric devices that are able to capture and utilize thermal energy from their cryogenic environment~\cite{PekolaHekking2007, Koski2014, Cottet2017, Brask2015, Rasola2024, Marchegiani2018, Aamir2023, Sundelin2024}. Building on existing device architectures enables reaching the state-of-the-art performance with the thermal powering scheme providing an advantage in energy efficiency.
    
In microwave circuits, the thermal energy of a system manifests itself as fluctuations of voltage across a resistive transmission line~\cite{Nyquist1928, Pozar2012}. Here we demonstrate that similar voltage fluctuations can provoke enhanced local dissipation in a superconducting circuit. The key part of our system is the quantum-circuit refrigerator (QCR)~\cite{Tan2017, Silveri2017, Vadimov2022}, which has previously exhibited versatility as a tunable environment by providing on-demand dissipation for superconducting resonators~\cite{Silveri2019, Sevriuk2019, Viitanen2021, Viitanen2024}, operating as an incoherent microwave photon source~\cite{Masuda2018}, demonstrating potential for realizing a quantum heat engine~\cite{Morstedt2024}, and enabling rapid initialization of superconducting qubits~\cite{Sevriuk2022, Yoshioka2023, Teixeira2024}. The dissipative effect of the QCR on its coupled system arises from incoherent absorption of photons in photon-assisted quasiparticle tunneling through a normal-metal--insulator--superconductor (NIS) junction. Typically, the system coupled to the QCR involves only a small population of relatively low-frequency photons compared to the size of the energy gap of the superconducting electrode. Thus, the QCR-induced rate of dissipation on the system can be tuned by several orders of magnitude by adjusting the bias voltage or coherent drive applied over the junction~\cite{Silveri2017, Vadimov2022, HsuSilveri2021,Viitanen2024}. However, the utilization of noise as a power source for the QCR has not been considered in the literature.

Owing to its broad frequency band of photon absorption enabling removal of multi-level excitations~\cite{Viitanen2021, Morstedt2022}, its ability to rapidly cool the coupled system to a low-temperature state independent of the initial state~\cite{Sevriuk2019, Yoshioka2023, Silveri2017}, and its straightforward integrability with arbitrary linear and nonlinear quantum-circuit elements~\cite{Hsu2020}, the QCR stands out among the recent experimental efforts~\cite{Aamir2023, Sundelin2024} in demonstrating an autonomous quantum refrigerator in a superconducting circuit. In addition to efficient reset, control, and readout applications in superconducting quantum-information systems, the prospects of the expanding group of thermally powered quantum devices may extend to autonomous timekeeping~\cite{Guzman2024, Erker2017}, quantum metrology~\cite{Bhattacharjee2021, Hofer2017}, and study of complex quantum thermodynamic processes~\cite{Myers2022}.

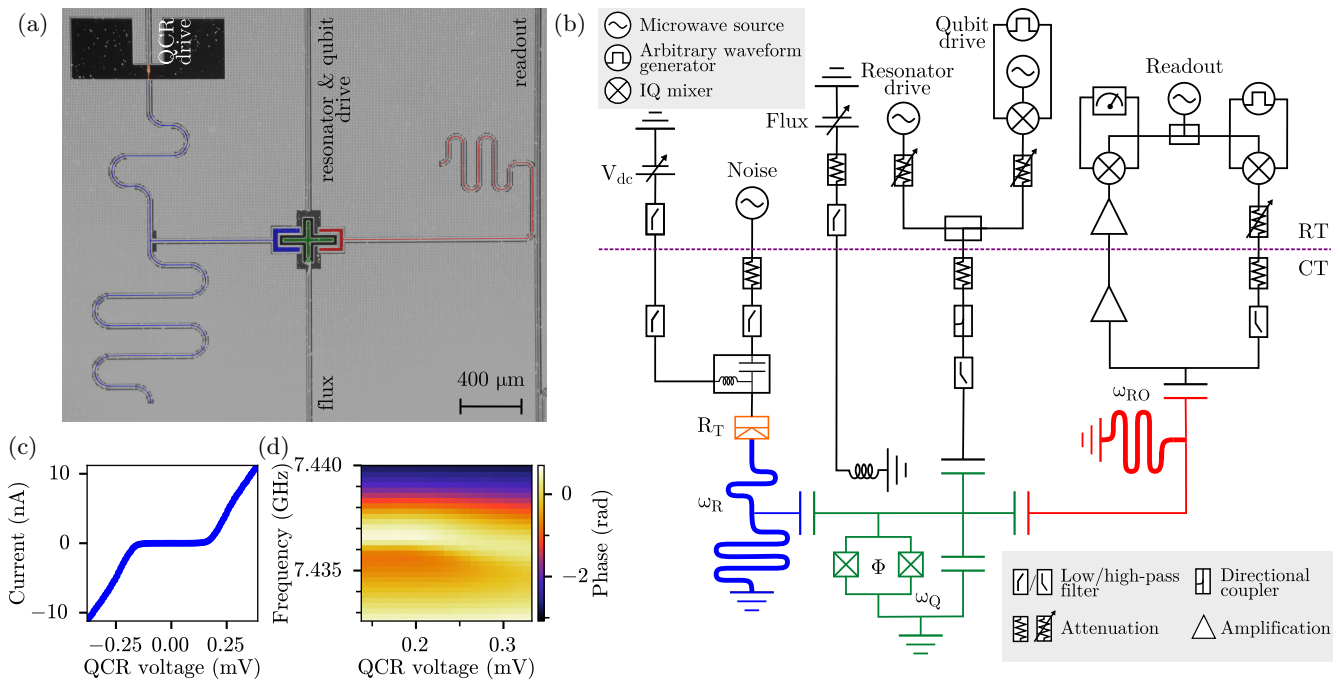
\begin{figure*}[]
    \centering
    \setlength{\unitlength}{1cm}
    \begin{minipage}[b!]{0.34 \textwidth}
        \subfloat{
            \begin{picture}(7.1,5.5)
                \put(0.5,0){\includegraphics{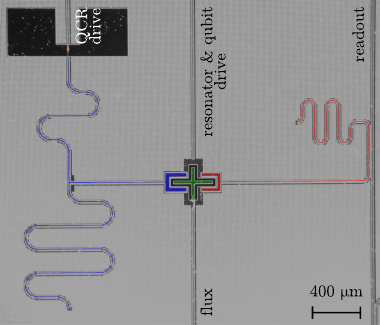}}
                \put(-0.1,5.2){(a)}
                \label{fig:sample_setup:sample_photo}
            \end{picture}
        }
    \end{minipage}
    \begin{minipage}[b!]{0.64 \textwidth}
        \subfloat{
            \begin{picture}(10,5.5)
                \put(0.8,-3.2){\includegraphics{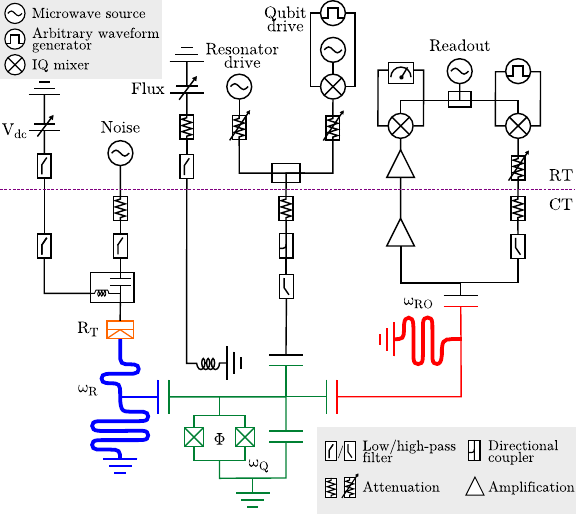}}
                \put(0.2,5.2){(b)}
                \label{fig:sample_setup:setup}
            \end{picture}
        }
    \end{minipage}\\
    \begin{minipage}[b!]{\textwidth}
            \subfloat{
            \begin{picture}(3,3.2)
                \put(-0.1,-0.2){\includegraphics{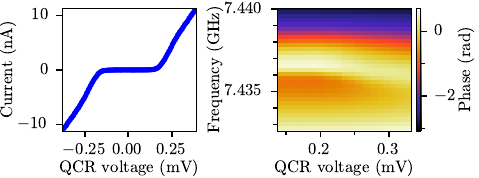}}
                \put(-0.1,2.8){(c)}
                \label{fig:sample_setup:IV}
            \end{picture}
        }
        \subfloat{
            \begin{picture}(\textwidth-3,0)
                \put(0,2.8){(d)}
                \label{fig:sample_setup:QCR_to_RO}
            \end{picture}
        }   
    \end{minipage}
    \caption{Device, experimental setup, and characteristics. (a)~False-color optical micrograph of a sample similar to that measured. The quarter-wave resonator (blue) is galvanically connected to the QCR (orange) from the superconductor side of the tunnel junction and probed with a transmon qubit (green) and its readout resonator (red). The device is controlled via the QCR driveline, the qubit--resonator driveline, the flux line, and the readout transmission line. (b)~Circuit diagram of the experimental setup. The colors of the components on the sample chip match to those in (a). The green crossed boxes denote the superconducting Josephson junctions of the qubit, whereas the orange half-crossed box illustrates the normal-metal and superconductor sides of the QCR junction. The boundary between the room-temperature (RT) and cryogenic-temperature (CT) components inside the dilution refrigerator is indicated with a purple dashed line. (c)~Current through the QCR junction as a function of the dc voltage applied across it without other driven excitations in the circuit. (d)~Phase of the readout signal as a function of QCR dc voltage and the frequency of the readout signal. A shift in the readout resonance frequency can be observed above the gap voltage $\Delta/e=203$~\textmu V which is extracted from (c). Here, $e$ is the elementary charge.}
    \label{fig:sample_setup}
\end{figure*} 

In this work, we demonstrate a distinct reduction in coherent and thermal populations of a superconducting resonator induced by a galvanically connected single-junction QCR driven purely by noise, with no dc voltage applied over the junction. This is achieved by generating a high-power noise signal from a room-temperature source and monitoring the population of the resonator with a transmon qubit probe~\cite{Schuster2007, Viitanen2024}. We find that an ac-drive model similar to Ref.~\cite{HsuSilveri2021} captures the essential characteristics of the noise-driven QCR by comparing the model with the measurement data. Our results pave the way for autonomous nanoelectric devices only powered by the thermal noise arising from their environments such as attenuators thermalized at the high-temperature stages of the cryostat.

\section{\label{sec:setup}Device and setup}

Our sample depicted in Fig.~\ref{fig:sample_setup:sample_photo} comprises a superconducting niobium quarter-wave coplanar-waveguide (CPW) resonator galvanically connected to an Al--Al$_2$O$_3$--Cu single-junction QCR from the side of the superconducting aluminum electrode. In addition, the resonator is capacitively coupled to a transmon qubit and its readout resonator patterned out of the same Nb layer. The resonant frequencies of the resonators and of the transmon given in Table~\ref{tab:sample_parameters} are well separated, enabling the detection of the resonator state in the strong-dispersive regime. The sample fabrication process is described in our previous publications~\cite{Viitanen2024,Teixeira2024,Morstedt2024}.

\begin{table}[]
    \centering
    \caption{Characteristic parameters of the sample and of the setup. Apart from the four bottommost quantities, the values are obtained experimentally. Here, $h$ is the Planck constant.}
    \begin{tabular}{l c c c c}
        \hline
         Parameter&& Symbol&& Value \\\hline
         Resonator frequency&& $\omega_{\mathrm{R}}/(2\pi)$&& 4.671 GHz\\
         Qubit frequency&& $\omega_{\mathrm{Q}}/(2\pi)$&& 3.953 GHz\\
         Readout resonator frequency&& $\omega_{\mathrm{RO}}/(2\pi)$&& 7.436 GHz\\
         Qubit anharmonicity&& $\alpha/(2\pi)$&& -275 MHz\\
         Resonator dispersive shift&& $\chi_{\mathrm{R}}/(2\pi)$&& 2.2 MHz\\
         Readout dispersive shift&& $\chi_{\mathrm{RO}}/(2\pi)$&& 0.6 MHz\\
         Qubit--resonator coupling strength&& $g_{\mathrm{R}}/(2\pi)$&& 76 MHz\\
         Qubit--readout coupling strength&& $g_{\mathrm{RO}}/(2\pi)$&& 169 MHz\\
         Gap frequency&& $\Delta/h$&& 49 GHz\\
         Driveline coupling strength&& $\gamma_{\mathrm{dr}}/(2\pi)$&& 1.1 MHz\\
         Excess coupling strength&& $\gamma_0/(2\pi)$&& 1.3 MHz\\
         Dynes parameter&& $\gamma_{\mathrm{D}}$&& $1.96\times 10^{-3}$\\
         Tunneling resistance&& $R_{\mathrm{T}}$&& 29.4 k$\mathrm{\Omega}$\\
         Junction capacitance&& $C_{\mathrm{NIS}}$&& 0.54 fF\\
         CPW characteristic impedance&& $Z_{\mathrm{0}}$&& 50 $\mathrm{\Omega}$\\
         AFM noise center frequency&& $\omega_{\mathrm{N,AFM}}/(2\pi)$&& 3.6 GHz\\
         VFM noise center frequency&& $\omega_{\mathrm{N,VFM}}/(2\pi)$&& 3.2 GHz\\\hline
    \end{tabular}
    \label{tab:sample_parameters}
\end{table}

The experimental setup is illustrated in Fig.~\ref{fig:sample_setup:setup}. The device is controlled via four CPW drivelines: the QCR driveline allows for both dc voltage and noise through a bias tee, the resonator and the qubit are driven via a shared driveline with separate sources and attenuation, the dispersive readout is carried out via the readout transmission line, and the qubit frequency is tuned via the dc-biased flux line. Figure~\ref{fig:sample_setup:IV} presents the dc current--voltage (IV) curve of the NIS junction without any additional drive signals applied at the input ports of the device. Here, we extract the superconductor gap parameter $\Delta\approx 203$~\textmu eV, the Dynes parameter $\gamma_{\mathrm{D}}\approx 1.96\times 10^{-3}$, and the tunneling resistance $R_{\mathrm{T}}\approx 29.4\ \mathrm{k\Omega}$ by fitting the elastic-tunneling model discussed in more detail in Sec.~\ref{sec:model}, assuming purely dc excitation. Note that elastic tunneling may be used to cool the quasiparticle reservoir in the normal metal~\cite{Giazotto2006, Zhang2015, PekolaGiazotto2007, Saira2007} whereas inelastic tunneling is at the heart of the QCR operation.

Figure~\ref{fig:sample_setup:QCR_to_RO} illustrates the effect of the dc-voltage-biased QCR on the readout resonator. Around $V_{\mathrm{dc}} = 230$~\textmu V bias voltage, the resonance frequency of the readout resonator shifts down by approximately 1 MHz. Since the shift occurs at bias voltages where the incoherent photon emission rate from the QCR is expected to be large, we conclude this shift to stem from the increased population in the transmon and the readout resonator by the QCR-emitted photons, changing the dispersive contribution of the qubit to the readout frequency~\cite{Masuda2018, Blais2021}. Consequently, in experiments that utilize a fixed-frequency readout tone, this effect results in a loss of the readout signal when the QCR is excited above the gap. The optimal operation regime of the QCR with negligible photon emission is slightly below the gap voltage, where the photons occupying the resonator are able to contribute to the quasiparticle tunneling by adding up to the energy required to reach the gap edge.

\section{\label{sec:noise_char}Noise characteristics}

\begin{figure*}[]
    \setlength{\unitlength}{1cm}
    \begin{minipage}[b!]{0.49 \textwidth}
            \addtocounter{subfigure}{1}
            \subfloat{
            \begin{picture}(9,6.5)
                \put(0,0){\includegraphics{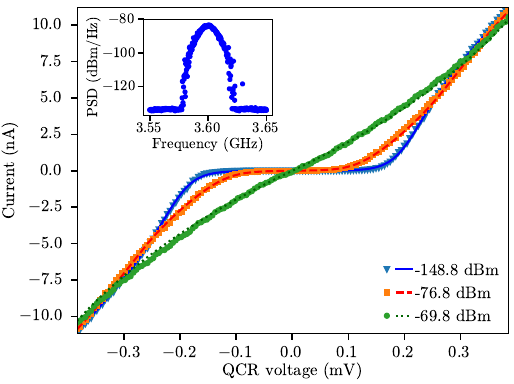}}
                \put(0,6){(b)}
                \label{fig:AFM:IV}
            \end{picture}
            }\\
            \addtocounter{subfigure}{-2}
            \subfloat{
            \begin{picture}(9,0)
                \put(2.5,6.2){(a)}
                \label{fig:AFM:power_spectrum}
            \end{picture}
            }
    \end{minipage}
    \begin{minipage}[b!]{0.49 \textwidth}
            \addtocounter{subfigure}{2}
            \subfloat{
            \begin{picture}(9,6.5)
                \put(0,0){\includegraphics{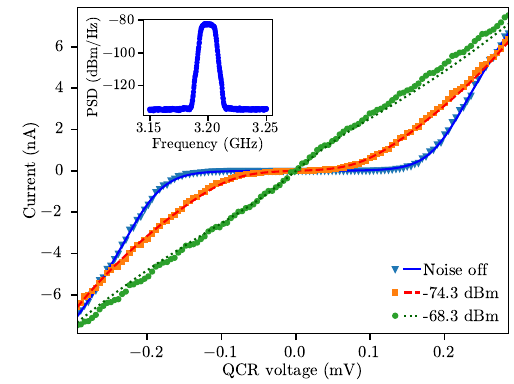}}
                \put(0,6){(d)}
                \label{fig:VFM:IV}
            \end{picture}
            }\\
            \addtocounter{subfigure}{-2}
            \subfloat{
            \begin{picture}(9,0)
                \put(2.5,6.2){(c)}
                \label{fig:VFM:power_spectrum}
            \end{picture}
            }
    \end{minipage}
    \caption{Noise spectra and its effect on the current--voltage characteristics of the QCR. (a)~Measured power spectral density of the AFM noise for a carrier frequency of 3.6~GHz and a nominal output power of $-9$~dBm. (b)~Measured (markers) and modeled (lines) current through the QCR junction as a function of dc bias voltage, with the AFM noise drive displayed in (a) applied to the QCR driveline at different powers incident to the QCR junction. We use the analytical model of Eq.~\eqref{eq:I} with parameters listed in Table~\ref{tab:sample_parameters} and the quasiparticle temperature of $T_{\mathrm{qp}}=248$~mK. (c)~Measured power spectral density of the VFM noise for a carrier frequency of 3.2~GHz and a nominal output power of $-10$~dBm. (d)~As (b)~but for the VFM noise and the analytical model of Eq.~\eqref{eq:I} uses the parameters of Table~\ref{tab:sample_parameters}, and quasiparticle temperature $T_{\mathrm{qp}}=250$~mK. The data in (a) and (c) are averaged over 100 traces with a resolution bandwidth of 20 kHz, and the data in (b) and (d) are averaged over 10 repetitions.}
    \label{fig:IVs_and_noise_spectra}
\end{figure*}
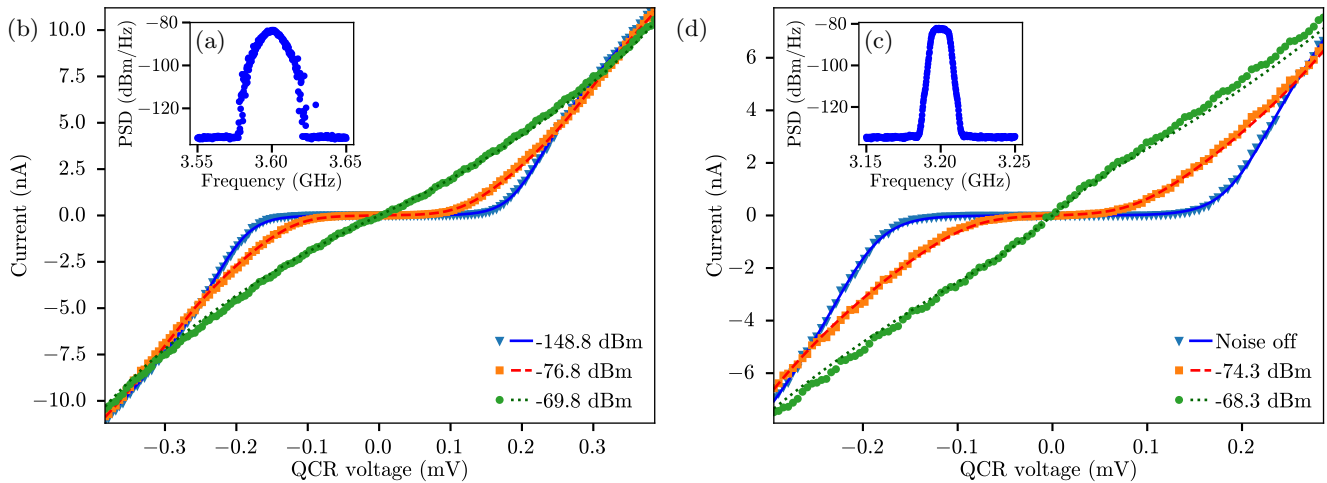

\begin{figure}[]
    \setlength{\unitlength}{1cm}
    \begin{minipage}[b!]{0.49\columnwidth}
            \subfloat{
            \begin{picture}(0.99\linewidth,6.5)
                \put(-0.2,0){\includegraphics{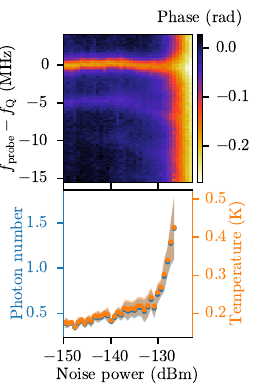}}
                \put(0.4,5.8){(a)}
                \label{fig:AFM:reso_q_spectrum}
            \end{picture}
            }\\
            \subfloat{
            \begin{picture}(0.99\linewidth,0)
                \put(0.4,3.4){(b)}
                \label{fig:AFM:reso_nT}
            \end{picture}
            }
    \end{minipage}
    \begin{minipage}[b!]{0.49\columnwidth}
            \subfloat{
            \begin{picture}(0.99\linewidth,6.5)
                \put(-0.1,0){\includegraphics{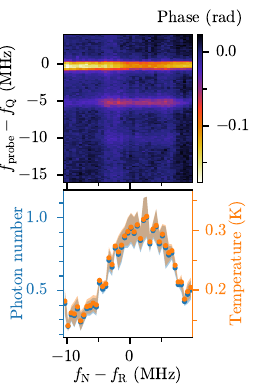}}
                \put(0.4,5.8){(c)}
                \label{fig:VFM:reso_q_spectrum}
            \end{picture}
            }\\
            \subfloat{
            \begin{picture}(0.99\linewidth,0)
                \put(0.4,3.4){(d)}
                \label{fig:VFM:reso_nT}
            \end{picture}
            }
    \end{minipage}
    \caption{Thermal state of a noise-driven resonator without QCR dc bias. (a)~Normalized phase of the readout signal as a function of the AFM noise power and the qubit probe detuning. The AFM noise of Fig.~\ref{fig:AFM:power_spectrum} with its center frequency tuned to match the resonator frequency $\omega_{\mathrm{R}}$ is applied to the qubit--resonator driveline. (b)~Mean photon number in the resonator (blue color, left axis) and the corresponding effective temperature of the resonator mode (orange color, right axis) as functions of AFM noise power. (c)~Normalized phase of the readout signal as a function of the detuning of the center frequency of the VFM noise from the resonator frequency and the qubit probe detuning. The VFM noise of Fig.~\ref{fig:VFM:power_spectrum} is applied to the qubit--resonator driveline at $-138.4$-dBm noise power incident to the resonator. (d)~Mean photon number in the resonator (blue color, left axis) and the corresponding effective temperature of the resonator mode (orange color, right axis) as functions of VFM noise detuning. The dots in (b) and (d) are the parameter values extracted from a fit of thermal distribution to the magnitude of the spectral lines in (a) and (c), respectively, with the shading denoting their $1\sigma$ confidence intervals owing to fit uncertainty.}
    \label{fig:VFM}
\end{figure}

The noise drive of the QCR is generated with an SRS SG396 signal generator utilizing two different signal modulation schemes. The first type of noise is obtained by analog frequency modulation (AFM), i.e., by modulating the frequency of the carrier signal with an analog pseudo-random additive white Gaussian noise waveform. The power spectrum corresponding to this AFM noiseform is shown in Fig.~\ref{fig:AFM:power_spectrum}. The second type of noise is obtained by vector frequency modulation (VFM), applying a frequency-shift keying scheme that utilizes a discrete pseudo-random binary sequence waveform to modulate the frequency of the carrier. The power spectrum of the resulting VFM noise is shown in Fig.~\ref{fig:VFM:power_spectrum}. Owing to the limitations of the source, both noise types have a relatively narrow 3-dB bandwidth of 12~MHz, the full bandwidth corresponding to approximately 46~MHz for the AFM noise and 31~MHz for the VFM noise. However, the scenario is very similar to a pure thermal noise drive bandlimited by on-chip filters such as an auxiliary resonator, where only a spectrum of frequencies which is narrow compared to the gap frequency arrives at the QCR. Unfiltered ideal thermal noise with photon energies exceeding the gap is not optimal for refrigeration since the high-energy photons will increase the rate of incoherent photon emission which eventually leads to quasiparticle poisoning.

The heating of the resonator by the AFM noise with the center frequency set to match the resonator frequency is presented in Figs.~\ref{fig:AFM:reso_q_spectrum} and~\ref{fig:AFM:reso_nT}. We monitor the resonator state by sweeping the qubit probe frequency near the ground-state--excited-state ($g$--$e$) transition frequency and measuring the transmitted signal through the readout line. Owing to the strong dispersive coupling between the resonator and the qubit, the Fock states of the resonator are distinguishable as equidistant resonance peaks in the qubit spectrum~\cite{Schuster2007,Viitanen2024}. By averaging over a large number of frequency sweeps, the relative heights of the spectral peaks follow the underlying probability distribution of the photon occupation in the resonator, from which the mean photon number can be extracted as a fitting parameter. Since the noise bandwidth is an order of magnitude greater than the resonator linewidth with the QCR in the unbiased off-state $\gamma(0)/(2\pi)=2.4$~MHz, the resonator population is expected to follow the thermal distribution $p_{\mathrm{t}}(n)=\Bar{n}^n/(\Bar{n}+1)^{n+1}$. By increasing the noise power, the mean thermal population of the resonator can be increased by more than a single photon, beyond a resonator temperature of 400~mK. 

At high noise power, a power-dependent increase in the spectral line splitting can be observed in Fig.~\ref{fig:AFM:reso_q_spectrum}, which we attribute to the nonlinear higher-order effects and broadening of the qubit linewidth introduced by the strong drive~\cite{Blais2021,Gambetta2006}. Furthermore, the strong noise drive activates the quasiparticle tunneling in the QCR via multiphoton absorption, ultimately inducing the observed qubit-state-independent background phase shift and loss of the readout signal as discussed in Sec.~\ref{sec:setup}.

The observed background shift in the absence of the QCR dc bias suggests that the QCR can be activated purely by the noise drive. However, to induce dissipation, the heating effect of the noise drive on the refrigerated mode has to be eliminated by filtering or, equivalently in the case of artificial noise, by tuning the noiseband away from the resonator frequency. Figures~\ref{fig:VFM:reso_q_spectrum} and~\ref{fig:VFM:reso_nT} show the effect of tuning the VFM noise drive on the thermal resonator population. Compared with a 10-MHz-detuned noise drive with negligible contribution to the resonator population, a resonant noise drive has a distinct increasing effect on the population. The slightly asymmetric character of the resonator population between the positive and negative detuning is attributed to the asymmetric resonance peak of the resonator (data not shown). By selecting noise frequencies $\omega_{\mathrm{N,AFM}}/(2\pi) =3.6$~GHz and $\omega_{\mathrm{N,VFM}}/(2\pi) =3.2$~GHz, which are far-detuned from the resonator, to drive the QCR, we expect the induced parasitic heating of the resonator to be negligible.

Figure~\ref{fig:AFM:IV} shows the IV curves of the NIS junction with an additional AFM noise drive applied to the QCR driveline at different levels of power. For sufficiently high power levels, a significant part of the quasiparticle energy required for tunneling can be obtained from the noise drive, visible as the suppression of the subgap current plateau. Around $-70$-dBm noise power, the plateau vanishes completely, indicating that the quasiparticle tunneling through the junction is enhanced by noise so strongly that the energy gap in the superconductor density of states does not suppress the current. Similar IV curves can be obtained with the VFM noise, as shown in Fig.~\ref{fig:VFM:IV}. 

\section{\label{sec:model} Model of noise-driven QCR coupled to resonator}

An approximate description of the effect of the noise-driven QCR is obtained by crudely replacing the finite-band noise drive by a sinusoidal drive at the center frequency of the noise $\omega_{\mathrm{ac}} = \omega_{\mathrm{N}}$. We define the total time-dependent voltage across the QCR junction as $V(t) = V_{\mathrm{dc}} + V_{\mathrm{ac}}\cos(\omega_{\mathrm{ac}}t)$, where $V_{\mathrm{dc}}$ is the dc part of the voltage, $V_{\mathrm{ac}}=2\sqrt{2P_{\mathrm{N}}Z_0}$ is the amplitude of the ac part given by the noise drive, $P_{\mathrm{N}}$ is the input noise power incident to the junction, and $Z_0$ is the characteristic impedance of the CPW. Here, the factor of two is the voltage transmission coefficient at the strongly mismatched CPW--junction interface. Following Ref.~\cite{HsuSilveri2021}, the ac component of the drive promotes quasiparticle tunneling events in addition to those present with pure dc bias. The resulting resonator transition rate from the Fock state $m$ to $m'$ associated with forward quasiparticle tunneling is given by
\begin{align}\label{eq:rate}
    \overrightarrow{\Gamma}_{mm'}(V) \approx\ &M^2_{mm'}\frac{R_{\mathrm{K}}}{R_{\mathrm{T}}}\sum_{k=-\infty}^{\infty}\Biggl\{\left[J_k\left(\frac{eV_{\mathrm{ac}}}{\hbar\omega_{\mathrm{ac}}}\right)\right]^2 \nonumber\\
    &\times \overrightarrow{F}\left[-eV_{\mathrm{dc}}+k\hbar\omega_{\mathrm{ac}}+\hbar\omega_{\mathrm{R}}(m-m')\right]\Biggr\},
\end{align}
where $M_{mm'}$ is the related matrix element, $R_{\mathrm{K}}$ is the von Klitzing constant, $J_k$ is the $k$th Bessel function of the first kind, and $\hbar$ is the reduced Planck constant. The normalized forward tunneling rate of quasiparticles $\overrightarrow{F}$ is defined as
\begin{equation}
    \overrightarrow{F}(E) = \frac{1}{h}\int\mathrm{d}\varepsilon\, n_{\mathrm{S}}(\varepsilon)\left[1-f_{\mathrm{S}}(\varepsilon)\right]f_{\mathrm{N}}(\varepsilon-E), 
\end{equation}
where $f_{\mathrm{S}}$ and $f_{\mathrm{N}}$ are the Fermi functions of the superconducting and normal-metal electrodes, and $n_{\mathrm{S}}$ is the normalized Dynes density of states~\cite{Silveri2017, Dynes1978}. Similarly, assuming identical temperatures on both electrodes, the transition rate associated with backward tunneling is obtained from $\overleftarrow{\Gamma}_{mm'}(V)=\overrightarrow{\Gamma}_{mm'}(-V)$. For a typical low-impedance CPW resonator, the inelastic transition rates with $m\neq m'$ are orders of magnitude smaller than the elastic transitions $\overrightarrow{\Gamma}_{mm}=\overrightarrow{\Gamma}_{00}$~\cite{Silveri2017}. Thus, we may neglect the contribution of inelastic processes to the tunneling current through the NIS junction and define 
\begin{equation}\label{eq:I}
    I(V) \approx -e\left[ \overrightarrow{\Gamma}_{00}(V)-\overleftarrow{\Gamma}_{00}(V)\right].
\end{equation}

Our analytical model of Eq.~\eqref{eq:I} agrees well with the data for both types of noise in Figs.~\ref{fig:AFM:IV} and\ref{fig:VFM:IV} . The small jumps distinguishable at the highest-power curves most likely arise from fluctuations of the ground potential interfering with the measurement.

\section{\label{sec:cooling_results}Noise-driven refrigeration of coherent and thermal states}

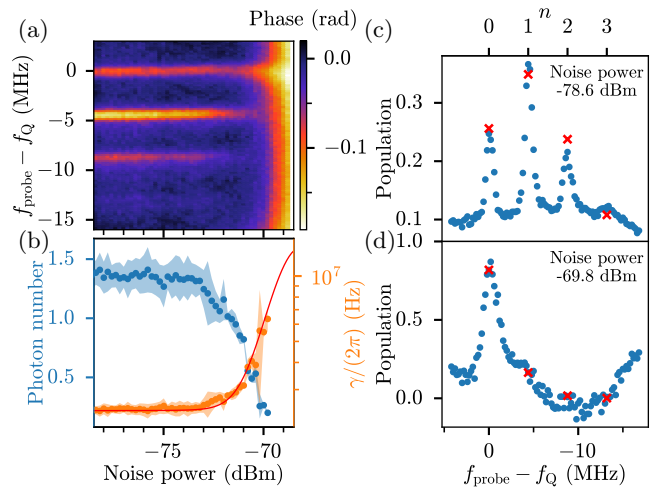
\begin{figure}[]
    \setlength{\unitlength}{1cm}
    \includegraphics{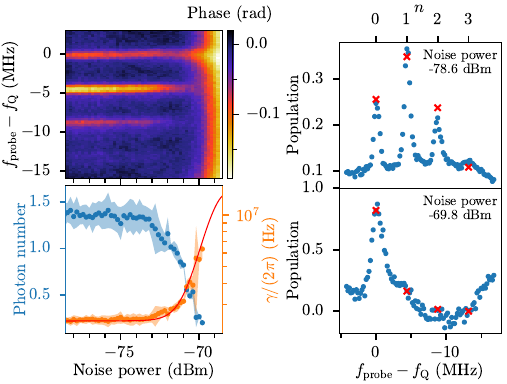}
    \subfloat{
        \label{fig:coherent:spectrum}
    }
    \subfloat{
        \label{fig:coherent:n_decay}
    }
    \subfloat{
        \label{fig:coherent:poisson_lopow}
    }
    \subfloat{
        \label{fig:coherent:poisson_hipow}
    }
    \begin{picture}(0,0)
        \put(-4.4,6.4){(a)}
        \put(-4.4,3.6){(b)}
        \put(0.2,6.4){(c)}
        \put(0.2,3.6){(d)}
    \end{picture}
    \caption{Damping of a coherent resonator state purely by noise. (a)~Normalized phase of the readout signal as a function of noise power applied to the QCR driveline and qubit probe detuning. The resonator is driven by an incident $-134.3$-dBm coherent tone at its resonance frequency. The QCR is driven by the AFM noise displayed in Fig.~\ref{fig:AFM:power_spectrum} without any dc bias voltage. (b) Mean photon number (blue color, left axis) and the corresponding decay rate (orange color, right axis) of the resonator as functions of the noise power. The blue dots are the parameter values extracted from a Poissonian fit to the magnitudes of the spectral lines in (a) and the orange dots are obtained by converting the extracted photon number to total decay rate utilizing Eqs.~\eqref{eq:power_balance} and~\eqref{eq:gamma} and the driveline coupling rate $\gamma_{\mathrm{dr}}/(2\pi)=1.1$~MHz. The shading denotes $1\sigma$ confidence intervals owing to fit uncertainty. The red line shows the decay rate given by the analytical model of Eq.~\eqref{eq:gamma_qcr} with the parameters of Table~\ref{tab:sample_parameters} and the quasiparticle temperature of $T_\mathrm{qp}=60$~mK. (c)~Scaled qubit spectrum from (a) (blue dots) and a least-square Poisson distribution fit (red crosses) with a very weak QCR drive. Each spectral peak corresponds to some Fock state $\ket{n}$, as indicated on the top axis. (d)~As (c), but with the QCR activated by a $-69.8$-dBm noise drive.}
    \label{fig:coherent}
\end{figure}

\begin{figure}[]
    \setlength{\unitlength}{1cm}
    \includegraphics{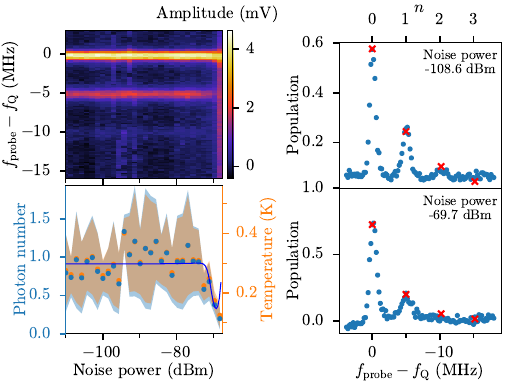}
    \subfloat{
        \label{fig:thermal:spectrum}
    }
    \subfloat{
        \label{fig:thermal:n_T}
    }
    \subfloat{
        \label{fig:thermal:lopow}
    }
    \subfloat{
        \label{fig:thermal:hipow}
    }
    \begin{picture}(0,0)
        \put(-4.4,6.4){(a)}
        \put(-4.4,3.6){(b)}
        \put(0.2,6.4){(c)}
        \put(0.2,3.6){(d)}
    \end{picture}
    \caption{Suppression of thermal resonator population purely by noise. (a)~Normalized amplitude of the readout signal as a function of noise power applied to the QCR driveline and qubit probe detuning. No drive is applied to the resonator. The QCR is driven with the VFM noise displayed in Fig.~\ref{fig:VFM:power_spectrum} without any dc bias voltage. (b)~Mean photon number (blue color, left axis) and the corresponding effective temperature (orange color, right axis) of the resonator as functions of noise power. The dots are the parameter values extracted from a thermal distribution fit to the magnitude of the spectral lines in (a) and the shading denotes their $1\sigma$ confidence intervals owing to fit uncertainty. The blue line shows the mean photon number given by the analytical model of Eq.~\eqref{eq:n} with the parameters of Table~\ref{tab:sample_parameters}, mean thermal population associated with QCR-independent coupling $n_{\mathrm{c}}=0.92$, and quasiparticle temperature $T_\mathrm{qp}=60$~mK. (c)~Scaled qubit spectrum from (a) (blue dots) and a least-square thermal distribution fit (red crosses) with a very weak QCR drive. Each spectral peak matches some Fock state $\ket{n}$, as indicated on the top axis. (d)~As (c), but with the QCR activated by a $-69.7$-dBm noise drive.}
    \label{fig:thermal}
\end{figure}
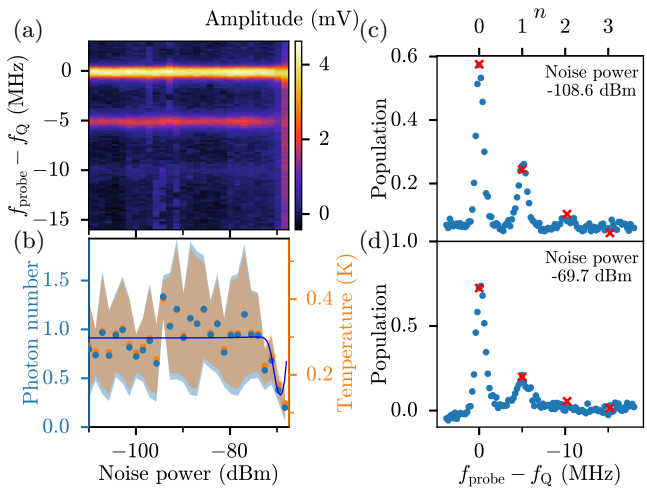

Figure~\ref{fig:coherent} presents the suppression of a coherent resonator state owing to the dissipation induced by a purely noise-driven QCR. Both the qubit spectrum in Fig.~\ref{fig:coherent:spectrum} and its fixed-power traces in Figs.~\ref{fig:coherent:poisson_lopow} and~\ref{fig:coherent:poisson_hipow} show a distinct shift in the relative heights of the spectral peaks with increased AFM noise power, consistent with a decreased mean photon number of the coherent state, as given by the Poisson distribution $p_{\mathrm{c}}(n) = e^{-n}\Bar{n}^n/n!$.

To compare the data with our analytical model, we first convert the extracted mean photon number to the total decay rate of the resonator $\gamma$ using the power balance
\begin{equation}\label{eq:power_balance}
    P_{\mathrm{in}}(1-|\Gamma_{\mathrm{dr}}|^2) = \hbar\omega_{\mathrm{R}}\Bar{n}(\gamma_{\mathrm{QCR}}+\gamma_0),
\end{equation}
where $P_{\mathrm{in}}$ is the power of the resonant drive in the driveline incident to the resonator, $\Gamma_{\mathrm{dr}} = \frac{\gamma_{\mathrm{dr}}-\gamma_{\mathrm{QCR}}-\gamma_0}{\gamma_{\mathrm{dr}}+\gamma_{\mathrm{QCR}}+\gamma_0}$ is the voltage reflection coefficient from the resonator, $\gamma_{\mathrm{dr}}$ is the decay rate to the driveline, $\gamma_{\mathrm{QCR}}$ is the QCR-induced decay rate, and $\gamma_0$ is the decay rate to excess sources. The total decay rate $\gamma$ is then obtained as the sum
\begin{equation}\label{eq:gamma}
    \gamma(V) = \gamma_{\mathrm{QCR}}(V) + \gamma_{\mathrm{c}},
\end{equation}
where $\gamma_{\mathrm{c}} = \gamma_{\mathrm{dr}} + \gamma_{\mathrm{0}}$ is the QCR-independent decay rate including the internal and external losses of the resonator. The analytical expression for the QCR-induced decay is given by the transition rates in Eq.~\eqref{eq:rate} as
\begin{equation}\label{eq:gamma_qcr}
    \gamma_{\mathrm{QCR}}(V) = \sum_{\tau=\pm 1}\left\{\overrightarrow{\Gamma}_{10}(\tau V)-\overrightarrow{\Gamma}_{01}(\tau V)\right\}.
\end{equation}

The extracted mean photon number and the decay rate of the coherently driven resonator is shown in Fig.~\ref{fig:coherent:n_decay}. Our analytical model of Eq.~\eqref{eq:gamma_qcr} is able to capture the noise-power-dependence of the resonator decay with a good accuracy. The strength of the dissipation provided by the purely noise-driven QCR agrees well with the previously demonstrated performance of dc- and rf-driven QCRs~\cite{Viitanen2024,Viitanen2021}.

The purely noise-driven cooling of a thermal resonator state is presented in Fig.~\ref{fig:thermal}. In the absence of a resonator drive, the mean photon number of the resonator is assumed to follow the weighted-average population of the coupled baths~\cite{Silveri2017,Viitanen2024}
\begin{equation}\label{eq:n}
    \Bar{n}(V)=\frac{n_{\mathrm{QCR}}(V)\gamma_{\mathrm{QCR}}(V)+n_{\mathrm{c}}\gamma_{\mathrm{c}}}{\gamma_{\mathrm{QCR}}(V)+\gamma_{\mathrm{c}}},
\end{equation}
where $n_{\mathrm{QCR}}$ is the bosonic population associated with the QCR bath, given by its effective temperature 
\begin{equation}
    T_{\mathrm{QCR}}(V) = \frac{\hbar\omega_{\mathrm{R}}}{k_{\mathrm{B}}}\left[\ln\left(\frac{\sum_{\tau=\pm1}\overrightarrow{\Gamma}_{10}(\tau V)}{\sum_{\tau=\pm1}\overrightarrow{\Gamma}_{01}(\tau V)}\right)\right]^{-1},
\end{equation}
$n_{\mathrm{c}}$ is the population associated with other losses in the resonator independent of the QCR, and $k_{\mathrm{B}}$ is the Boltzmann constant.

Figure~\ref{fig:thermal:n_T} shows the suppression of the thermal population in the resonator with increasing VFM noise power. Correspondingly, we observe that the noise applied on the QCR can reduce the effective temperature of the resonator mode roughly from 300~mK to 130~mK. The analytical model of Eq.~\eqref{eq:n} agrees with the data up to the fit uncertainty of the thermal distribution. Note that the fit uncertainty is relatively large at higher photon numbers owing to the monotonic shape of the thermal distribution. However, the visible change in the relative spectral line and peak magnitudes and the model following the trend of the data strongly support our observations.

\section{\label{sec:conclusions}Conclusions}

In conclusion, we have experimentally demonstrated on-demand suppression of the mean photon number in both thermal and coherent states of a coplanar-waveguide resonator by a fully noise-powered single-junction QCR, achieving a very good agreement between the data and our theoretical model. Since the same noise drive induces an elevated-temperature thermal state if applied to the resonator in resonance, we conclude that our noise drive corresponds to filtered thermal voltage fluctuations of a resistor or any other well-behaving dissipative environment. Although our current setup does not support driving the QCR with a passive thermal source owing to the high noise power required to activate the QCR, our results suggest that, as a first step, the requirements on the noise power can be significantly relieved by driving a resonator coupled to the QCR, in comparison to off-resonant driving at the QCR input. For further reduction of the required noise power in the future, we aim to enhance the multiphoton-assisted tunneling in the QCR by increasing the resonator impedance and mode frequency. Importantly, our results establish a strong foundation for single-junction-based autonomous in-situ refrigeration of superconducting circuits, enabling high energy efficiency and possible reliefs in scaling the future superconducting quantum processors. In the future, we aim to use a thermally powered QCR to implement an autonomous quantum heat engine~\cite{Morstedt2024}.

\begin{acknowledgments}
 This work was funded by the Academy of Finland Centre of Excellence program (project Nos. 352925 and 336810) and grant Nos.~316619 and 349594 (THEPOW), the European Research Council under Advanced Grant No.~101053801 (ConceptQ), and the Research Council of Finland through the Quantum Doctoral Education Pilot (QDOC) and the Finnish Quantum Flagship project (No.~358877, Aalto). We also acknowledge the provision of facilities and technical support by Aalto University at OtaNano—Micronova Nanofabrication Centre. We thank Wallace Teixeira, Vasilii Vadimov, Jukka Pekola, Bayan Karimi, Miika Rasola, Tuomas Uusnäkki, Ilari M\"akinen, Christoforus Satrya, Gianluigi Catelani, Joachim Ankerhold, J\"urgen Stockburger, Tapio Ala-Nissila, Andrea Trioni, and Johannes Fink for scientific discussions.
\end{acknowledgments}

\bibliographystyle{apsrev4-2}
\bibliography{noise_drive_article}

\end{document}